\documentclass{article}
\usepackage[utf8]{inputenc}
\usepackage[margin=1in]{geometry}
\usepackage[bookmarks=false]{hyperref}
\usepackage{graphicx}
\usepackage{booktabs}
\usepackage{authblk}
\usepackage[font=small,labelfont=bf]{caption}
\usepackage{subcaption}
\usepackage{enumitem}
\usepackage{algpseudocode}
\usepackage[ruled]{algorithm}
\title{Electroencephalogram (EEG) for Delineating Objective Measure of Autism Spectrum Disorder (ASD)}
\author[1]{Yasith Jayawardana}
\author[2]{Mark Jaime}
\author[3]{Sashi Thapaliya}
\author[1]{Sampath Jayarathna}
\affil[1]{Department of Computer Science, Old Dominion University, Norfolk, VA 23529, USA}
\affil[2]{Division of Science, Indiana University-Purdue University Columbus, IN 47203, USA}
\affil[3]{Department of Computer Science, California State Polytechnic University, Pomona, CA 91768, USA}
\date{}
\begin{document}

\maketitle

\begin{abstract}
Autism Spectrum Disorder (ASD) is a developmental disorder that often impairs a child’s normal development of the brain.
According to CDC, it is estimated that 1 in 6 children in the US suffer from development disorders, and 1 in 68 children in the US suffer from ASD.
This condition has a negative impact on a person’s ability to hear, socialize and communicate.
Overall, ASD has a broad range of symptoms and severity; hence the term spectrum is used.
One of the main contributors to ASD is known to be genetics.
Up to date, no suitable cure for ASD has been found.
Early diagnosis is crucial for the long-term treatment of ASD, but this is challenging due to the lack of a proper objective measures.
Subjective measures often take more time, resources, and have false positives or false negatives.
There is a need for efficient objective measures that can help in diagnosing this disease early as possible with less effort.

EEG measures the electric signals of the brain via electrodes placed on various places on the scalp.
These signals can be used to study complex neuropsychiatric issues.
Studies have shown that EEG has the potential to be used as a biomarker for various neurological conditions including ASD.
This chapter will outline the usage of EEG measurement for the classification of ASD using machine learning algorithms.
\end{abstract}

\section{Introduction}
Autism Spectrum Disorder (ASD) is characterized by significant impairments in social and communicative functioning as well as the presence of repetitive behaviors and/or restricted interests.
According to CDC estimates, the prevalence of ASD (14.6 per 1,000 children) has nearly doubled over the last decade and has a costly impact on the lives of families affected by the disorder.
It is estimated that 1 in 6 children in the US suffer from developmental disorders and 1 in 68 children fall under Autism Spectrum Disorder.
ASD is a neurological and developmental disorder that has negative impact in a person’s learning, social interaction and communication.
It is a debilitating condition that affects brain development from early childhood creating a lifelong challenge in normal functioning.
Autism is measured in spectrum because of the wide range of symptoms and severity. 
The total lifetime cost of care for an individual with ASD can be as high as \$2.4 million \cite{buescher2014costs}.
In the U.S., the long-term societal costs are projected to reach \$461 billion by 2025 \cite{leigh2015brief}. 

One of the main contributing factor for ASD is known to be genetics.
So far, no suitable cure has been found.
However, early intervention has been shown to reverse or correct most of its symptoms \cite{dawson2008early}.
And this can only be possible by early diagnosis.
Therefore, early diagnosis is crucial for successful treatment of ASD.
Although progress has been made to accurately diagnose ASD, it is far from ideal.
It often requires various tests such as behavioral assessments, observations from caretakers over a period of time to correctly determine the existence of Autism.
Even with this tedious testing often times individuals are misdiagnosed.
However, there remains promise in the development of accurate detection using various modalities of Biomedical Images, EEG, and Eye movements.

Efforts to identify feasible, low-cost, and etiologically meaningful
biobehavioral markers of ASD are thus critical for mitigating these costs through improvement in the objective detection of ASD.
However, the phenotypic and genotypic heterogeneity of ASD presents a unique challenge for identifying precursors aligned with currently recognized social processing dimensions of ASD.
One approach to unraveling the heterogeneity of ASD is to develop neurocognitive measures with shared coherence that map onto valid diagnostic tasks, like the Autism Diagnostic Observation Schedule Second Edition (ADOS-2) \cite{gotham2007autism}, that are the gold standard in ASD identification.
These measures can then be used to stratify children into homogeneous subgroups, each representing varying degrees of impaired social neurocognitive functioning.
Despite the need for objective, physiological measures of social functioning, machine learning has not yet been widely applied to biobehavioral metrics for diagnostic purposes in children with ASD.
 
This chapter focuses on a particular social processing domain which, according to the NIMH Research Domain Criteria (RDoC), is a central deficit of ASD and lends itself to quantifiable neurocognitive patterns: social interactions during ADOS-2. 
The ability to socially coordinate visual attention, share a point of view with another person, and process self- and other-related information
\cite{barresi1996intentional, butterworth1991minds, mundy2009parallel} is a foundational social cognitive capacity \cite{mundy2016autism}.
Its emergence in infancy predicts individual differences in language development in both children with ASD and in typically developing children \cite{mundy1990longitudinal, mundy2007attention}.
Moreover, attention is recognized in the diagnostic criteria of the DSM-V as one of the central impairments of early, nonverbal social communication in ASD.
While the empirical evidence on the physiological nature of attention deficits in ASD is emerging that can index attention: social brain functional connectivity (FC) during real-life social interaction.

At the same time, it is well-established in the literature that the neural systems that subserve social cognition are functionally compromised in children with ASD \cite{baron1985does, lombardo2011specialization, hill2003understanding, kana2009atypical, mason2008theory}.
The research suggests there is a functional (frontal-temporal-parietal) overlap in neural system activity during ADOS-2 and social cognitive processing \cite{mundy2016autism, kennedy2012social, redcay2012look, schurz2014fractionating, lombardo2010shared, caruana2015frontotemporoparietal}.
Taken together, there is ample evidence to support that aberrant frontal-temporal-parietal FC is a potential nexus for latent social cognitive disturbance in early ASD.

Many studies reveal either under- or over-connected areas in the autistic brain,
depending on whether the subject is at rest or engaged in cognitive processing \cite{coben2008eeg, just2004cortical, just2006functional, kana2014brain, koshino2005functional, koshino2007fmri, lazarev2015reduced, lynch2013default, uddin2013salience, shih2010atypical, noonan2009aberrant, jones2010sources, damarla2010cortical, mohammad2016brain}.
Reduced FC within frontal, superior--temporal, and temporal--parietal regions (regions that comprise the \emph{social brain system}) have been consistently reported in the majority of fMRI studies examining FC during social information processing \cite{koshino2007fmri, castelli2002autism, kleinhans2008abnormal, rudie2011reduced, welchew2005functional}. 
The presence of altered social brain system FC in early neurodevelopment can potentially reveal the onset of social difficulties \cite{keehn2013atypical}, as altered FC disrupts efficient information flow between parallel and distributed neural systems involved in the processing of social and communicative information \cite{mundy2009parallel}.
Thus, children with ASD may develop with limited neurocognitive resources to efficiently deal with the processing demands of dynamic social exchanges.
This social deficit may emerge as idiosyncratic patterns of EEG during bouts of joint social attention.

\section{Literature Survey}
\subsection{Social Interaction Tasks}
To date, the few studies that have examined FC during attention have done so using non-clinical paradigms that involve the observation of attention-eliciting videos; however, data from such paradigms may not reflect the true person-to-person interactive nature.
More importantly, video paradigms may only tap into one of two facets of attension: responding to joint attention (RJA), which serves an imperative function.
What is not represented in JA-eliciting video paradigms is initiating joint attention (IJA), which serves a declarative function and taps into social reward systems that are integral to the social sharing of experiences \cite{caruana2015frontotemporoparietal, schilbach2010minds,gordon2013social}.
Moreover, RJA and IJA show a developmental dissociation during the first and second years of life \cite{yoder2009predicting,ibanez2013development,mundy2007individual}.
Although RJA and IJA both have predictive value in infancy, IJA is a more stable marker of ASD than RJA in later childhood \cite{mundy1986defining}.
Some neuroimaging researchers have dealt with the above issues by using a live face-to-avatar paradigm to simulate IJA bids \cite{redcay2012look,gordon2013social}.
However, the movement constraints inside the MRI scanner create testing conditions that can be difficult for younger children, with and without ASD. 

Eye movement behavior is a result of complex neurological processes; therefore,  eye gaze metrics can reveal objective and quantifiable information about the predictability and consistency of covert social cognitive processes, including social attention \cite{chita-tegmark_social_2016, guillon_visual_2014}, emotion recognition \cite{bal_emotion_2010, black_mechanisms_2017, sawyer_can_2012, sasson_context_2016, tsang_eye-tracking_2018, wagner_greater_nodate, wieckowski_eye-gaze_2017}, perspective taking, \cite{symeonidou_development_2016} and joint attention \cite{bedford_precursors_2012, billeci_disentangling_2016, falck-ytter_gaze_2012, falck-ytter_brief_2015, swanson_broad_2013, thorup_altered_2016, thorup_reduced_2018, vivanti_social_2017} for children with and without ASD.
Eye gaze measurement includes a number of metrics relevant to oculomotor control \cite{komogortsev20132d} such as saccadic trajectories, fixations, and other relevant measures such as velocity, duration, amplitude, and pupil dilation \cite{krejtz2018eye}.
We believe that combined analysis of fixations and saccades during natural and dynamic joint attention tasks, currently used as a reliable measure of ASD diagnostic criteria, will represent valid biomarkers for objectifying and delineating the dimensionality of ASD diagnosis.
Previous work in this area have successfully demonstrated development of the coefficient of ambient/focal attention \cite{KDKS+16} and previous work has supported the relationship between eye tracking metrics and severity of ASD diagnosis \cite{frazier_development_2018, del_valle_rubido_search_2018} and communicative competence \cite{norbury_eye-movement_nodate}.

If visual attention influences stability of fixations dependant upon the demands of dynamic joint attention tasks, a natural next step is to look into how relevance may be reflected in eye movements and neurophysiologic features for atypical social brain systems, such as in the context of ASD \cite{hotier_social_nodate}.

\subsection{EEG based Machine Learning for ASD}
Studies have shown that EEG has the potential to be used as biomarker for various neurological conditions including ASD \cite{wang2013resting}.
EEG measures the electrical signals of the brain via electrodes that are placed on various places on the scalp.
These electrical signals are postsynaptic activity in the neocortex and can be used to study complex neuropsychiatric issues.
EEG has various frequency bands and its analysis are performed on these varying bandwidths.
Waves between 0.5 and 4 HZ are delta, between 4 and 8 HZ are theta, between 8 and 13 HZ are alpha, 13 to 35 HZ are beta and over 35 are gamma.

Saccadic eye movement plays a big role in the attention and behavior of an individual which directly affects both language and social skills \cite{fletcher2009eye}.
Autistic children seem to have different eye movement behaviors than non-autistic children.
They tend to avoid eye contact and looking at human face while focusing more on geometric shapes \cite{klin2009two}.
While a typical child doesn’t find any interest in geometric shapes and tend to make more eye contact, and human face perception.

In \cite{grossi2017diagnosis}, authors use a complex EEG processing algorithm called MSROM/I-FAST along with multiple machine learning algorithms to classify Autistic patients.
In this study 15 ASD individuals and 10 non ASD were selected.
ASD group comprised of 13 males and 2 females between 7 and 14 years of age.
Control group comprised of 4 males and 6 females between 7 and 12 years of age.
Resting State EEG of both closed and open eyes were recorded using 19 electrodes.
Patients sat in a quiet room without speaking or performing any mentally demanding activity while the EEG was being recorded.
The proposed IFAST algorithm consists of exactly three different phases or parts.
In the first stage also called Squashing phase, the raw EEG signals are converted into feature vectors.
Authors present a workflow of the system from raw data to classification to make comparison between different algorithms such as Multi Scale Entropy (MSE) and the Multi Scale Ranked Organizing Maps (MS-ROM).
MSROM is a novel algorithm based on Single Organizing Map Neural Network.
In this study, the dataset is randomly divided into 17 training consisting of 11 ASD, 6 controls and eight test records consisting of 4 ASD, 4 control.
The noise elimination is performed only on the training set.
Also it completely depends on the algorithm selected for extraction of feature vectors.
For MS-ROM features they utilize an algorithm called TWIST.
In the final classification stage they use multiple machine learning algorithms along with multiple validation protocols.
The validation protocols are training-testing and leave one out cross validation.
For classification purposes they make use of Sine Net Neural Network, Logistic Regression, Sequential Minimal Optimization, kNN, K-Contractive Map, Naive Bayes, and Random forest.
With MSE feature extraction the best results were given by Logistic and Naive Bayes with exactly 2 errors.
Whereas, MS-ROM with training test protocol had 0 errors (100\% accuracy) with all the classification models.

\cite{bosl2011eeg} conducts a study using mMSE as feature vectors along with multiclass Support Vector Machine to differentiate developing and high risk infant groups.
In this study they use 79 different infants of which 49 were considered high risk and 33 typically developing infants.
The 49 infants were high risk based on one of their older siblings having a confirmed ASD diagnosis.
The other 39 infants were not high risk based on the fact that no one in their family ever was diagnosed with ASD.
Data was collected from each infant during multiple sessions with some interval.
Data extracted from an infant in five different sessions in various months between 6 to 24 month period were considered unique.
Resting state EEG with 64 electrodes was extracted by placing the infant in a dimly lit room in their mother’s lap where the research assistant blew bubbles to catch their attention.
The raw signals were preprocessed using Modified MultiScale Entropy.
Low, high, and mean for each curve from mMSE were calculated to create a feature set of 192 values.
The best fit for the classification for High risk and normal infants was at age 9 months with over 90\% accuracy.

\cite{abdulhay2017frequency} uses EEG intrinsic function pulsation to identify patterns in Autism.
They mathematically compute EEG features and compare ASD with typically developing.
In this study they selected 10 children with ASD and 10 non autistic children within the age group of 4 to 13.
They collected resting state EEG using 64 electrodes with a 500 HZ sampling frequency.
Initially the signals were band pass filtered and all the artifacts including eye movements were removed by using Independent Component Analysis.
Empirical Mode decomposition was applied to extract Intrinsic Mode Function from each of the channels of the participants.
Then point by point pulsations of analytic intrinsic modes are computed which is then plotted to make comparison with the counterpart intrinsic mode in another channel.
Any existing stability loops are analyzed for abnormal neural connectivity.
In addition they perform 3D mapping to visualize and spot unusual brain activities.
In the first IMF of channel 3 versus the first IMF in channel 2 for typically developing and autistic child it was found that the stability of local pulsation pathways maintained consistency while it was random in typically developing.
Similar patterns were seen in channels 1 and 2 and 36 and 37 of non-autistic and autistic children.
Overall this computational method was able to differentiate the abnormal EEG activities between ASD and typically developing children.

\cite{alie2011analysis} uses Markov Models with eye tracking to classify Autism Spectrum Disorder.
Unlike most other studies that collected data from children who were 3 years or older, in this study they collect data from 6 month old infants.
There were in total 32 subjects out of which 6 were later at 3 years of age diagnosed with ASD and the rest were not.
During the data collection the subjects were placed in front of their mothers and four different cameras from different angles recorded the video for about 3 minutes.
The eye tracking was simply based on either the subject looked at the mother’s face or not.
Through this they get a binary sequence of subjects eye pattern which is then converted into alphabet sequence of a specific length.
Then the sequence was filtered using a low pass filter and down sampled by factor of 18.
This is done to enhance Markov Models to produce effective results. Using this data, they compare Hidden Markov Models and Variable-order Markov Models for the classification of ASD. Hidden Markov Models was able to correctly identified 92.03\% of the typically developing subject while identifying only 33.33\% of Autistic subject.
Whereas the VMM correctly identified 100\% of the Autistic and 92.03\% of typically developing subjects.
It was clear from this result that Variable-order markov models are superior in finding Autistic eye pattern while both Markov Models are the same in finding typically developing.
The authors point out this difference as a result of various spectrums of Autism with different eye patterns.
Nevertheless the VMM algorithm used in this study looks effective in identifying Autism in an early age.

Similarly, \cite{liu2015efficient} proposes a machine learning framework for the diagnosis of Autism using eye movement. 
They utilize two different datasets from previous studies.
One of the dataset had 20 ASD children, 21 typically developing, and 20 typical developing IQ-matched children.
The other dataset comprised of 19 ASD, 22 Intellectually disabled, and 28 typical young adults and adolescents.
They compute Bag of Words for Eye Coordinates and Eye movement, N-Grams and AOI from the datasets.
And they train five different Support Vector machine model with RBF kernel.
Each of the model used different form of features like BOW of eye coordinates, BOW of eye movement, combination, N-Grams, and AOI.
The end result was good for both groups with Combination or fusion data.
However, the children dataset with fusion was the best with around 87\% accuracy.

Another study \cite{jiang2017learning} uses eye movement with deep neural networks to identify individuals with Autism Spectrum Disorder.
They used dataset from a previous study with 20 ASD and 19 health controls.
Here the subjects observed around 700 images from the OSIE database.
OSIE database is a popular eye tracking dataset used for image saliency benchmarking.
First they use Cluster Fix algorithm on the raw data to compute fixations and saccades.
Next, they work on finding the discriminative images as the OSIE dataset is not specifically built for autism studies.
So, both groups might have the same visual pattern for some of the images.
For this purpose they use Fisher score method by which they score each of the images and select only the one with the higher scores to be processed further.
After this process of image selection they compute fixation maps in order to differentiate fixations between two groups.
Fixation maps are simply a probability distribution of all the eye fixations.
In addition they use a Gaussian Kernel for smoothing and normalize by their sum.
Normalization is usually done when we are comparing two different fixation maps as is the case here.
Then they compute difference of fixation map between the Autistic and non-Autistic group.
This is the original target which they used to train a SALICON network to predict these values.
SALICON network is one of the state of the art image saliency prediction algorithm.
Image saliency prediction is about predicting the visual pattern of users given an image.
SALICON network uses two VGG with 16 layers.
One of the VGG uses the original image to detect the small salient regions whereas the other VGG uses the down sampled image to detect the center of large salient regions.
At the end both the outputs are combined to get a better result.
This only predicts the image saliency.
So in order to predict the difference of fixation map they add another convolution layer with Cross Entropy Loss function using the original Difference of fixation map.
Next, they send the predicted difference of fixation maps to the final prediction layer.
In this part they first apply tanh function to the features then concatenate the feature vectors of all fixation in order to consider dynamic change of attention.
After which they reduce the dimension by using local average pooling.
At last they train a SVM to make the final classification between ASD and control.
They make use of the popular leave-one-out cross validation to measure the performance of their model.
The accuracy of this model showed real promise in eye tracking for ASD with about 92\% accuracy.

\section{Methodology}
Current techniques in practice for identifying ASD are mostly subjective and prone to error and usually takes a lot of time for final diagnosis.
Most of the children with ASD are diagnosed after 3 years of age.
Early diagnosis is the key for reversing or treating ASD through early intervention.
As time is of an essence we need a method of diagnosis that is fast, and efficient unlike the current practice that could take months to years.
Medical Imaging and blood testing \cite{sparks2002brain,spence2004autism} are promising and a lot of work is being done with these modalities to diagnose ASD.
However, EEG and Eye movement are cost effective and hence can be accessible in consumer level.

The aim of this research is to study the identification of Autism Spectrum Disorder using EEG during ADOS-2.
Comparison of the classification performance between EEG features can potentially result in finding the better feature set.
We hypothesize that the top performing signal most likely has more of the unique data points and pattern of ASD and similarly, the least performing signals have less of the data points and patterns relating to ASD. 
The secondary goal is to compare various machine learning algorithms for the classification purposes.
Conditions like ADHD, and other learning disabilities can also share similar comparative patterns for different features.

\subsection{EEG Acquisition and Pre-processing}
\label{sec:eeg-acq-pre-prop}
Our preliminary FC measures were analyzed from each pilot subject's EEG recording, acquired throughout the entire duration of the ADOS-2.
% ==========
\begin{figure}[hbt!]
    \centering
    \includegraphics[width=.8\textwidth]{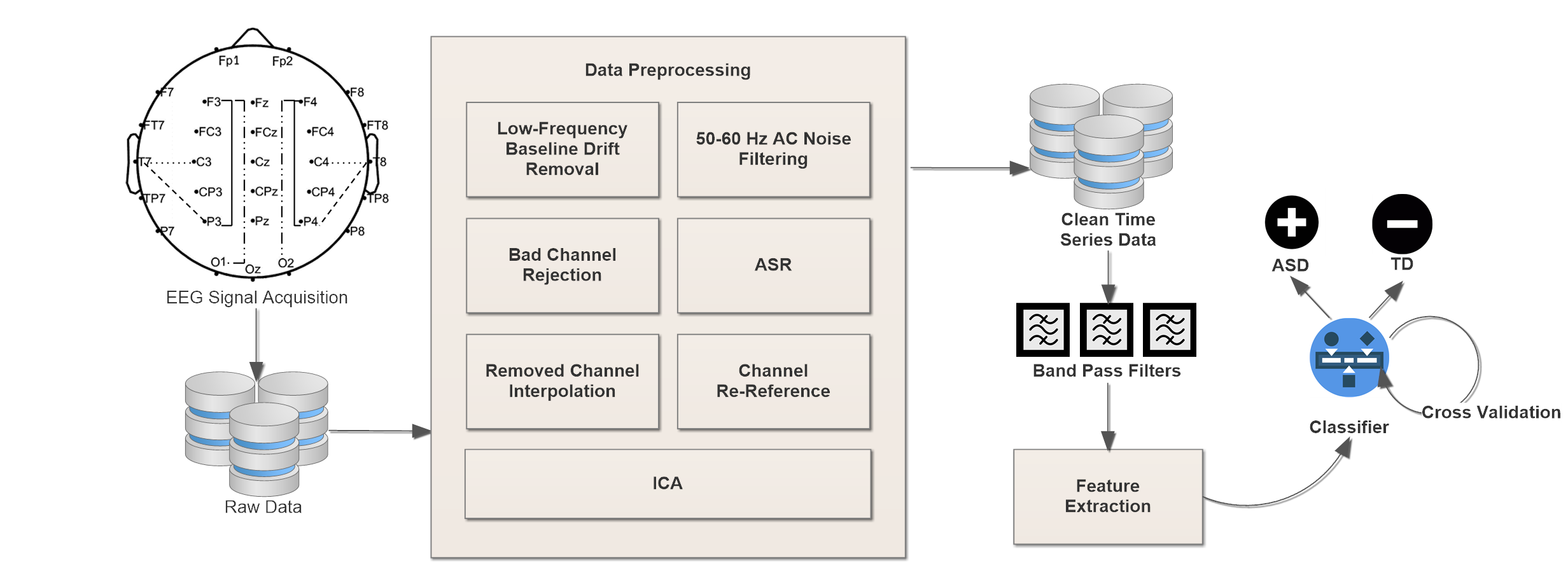}
    \caption{EEG Processing and Classification Pipeline}
    \label{fig:pp_pipeline}
\end{figure}
% ==========
We used a 32-channel LiveAmp wireless EEG system with active electrodes and a digital sampling rate of 250 Hz (Brain Products GmbH) for EEG time series acquisition.
Use of a wireless EEG system allowed for head movements and the active electrodes increased speed of application thereby increasing probability of successful EEG data acquisition with special populations.
All 32 channels were continuously recorded using the FCz electrode as reference.
To maximize the consistency of the recording quality across conditions, a single epoch was recorded per experimental condition.
In between epoch recordings an impedance check will be performed.
This was resulted in 6 different epochs per subject.
Prior to the recording of each experimental epoch, a 90 second epoch of eyes closed while resting will be recorded.
This will serve as a necessary baseline metric for the EEG analysis.

\begin{figure}[hbt!]
    \centering
    \begin{subfigure}[b]{0.45\textwidth}
        \centering
        \includegraphics[width=\textwidth]{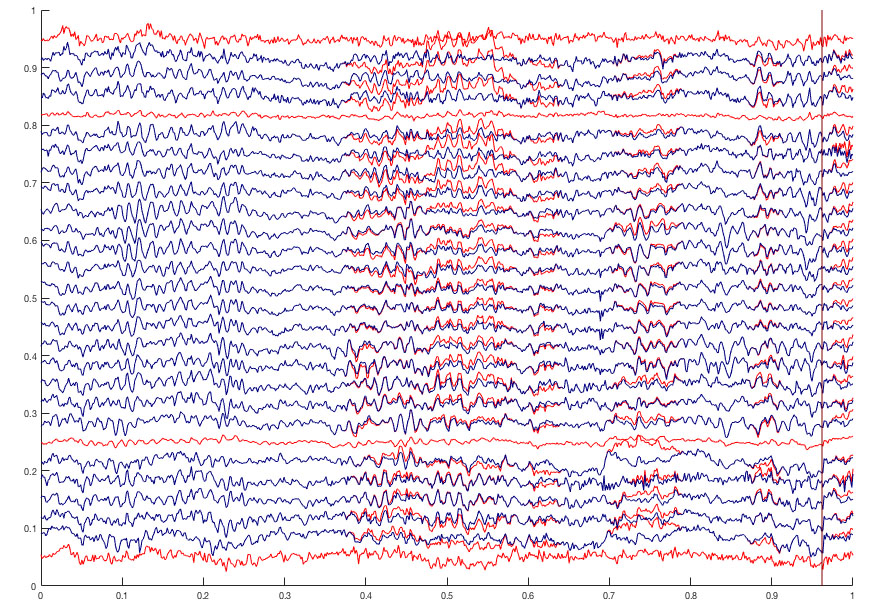}
        \caption{EEG time series.}
        \label{fig:eeg_ts}
    \end{subfigure}
    \hfill
    \begin{subfigure}[b]{0.45\textwidth}
        \centering
        \includegraphics[width=\textwidth]{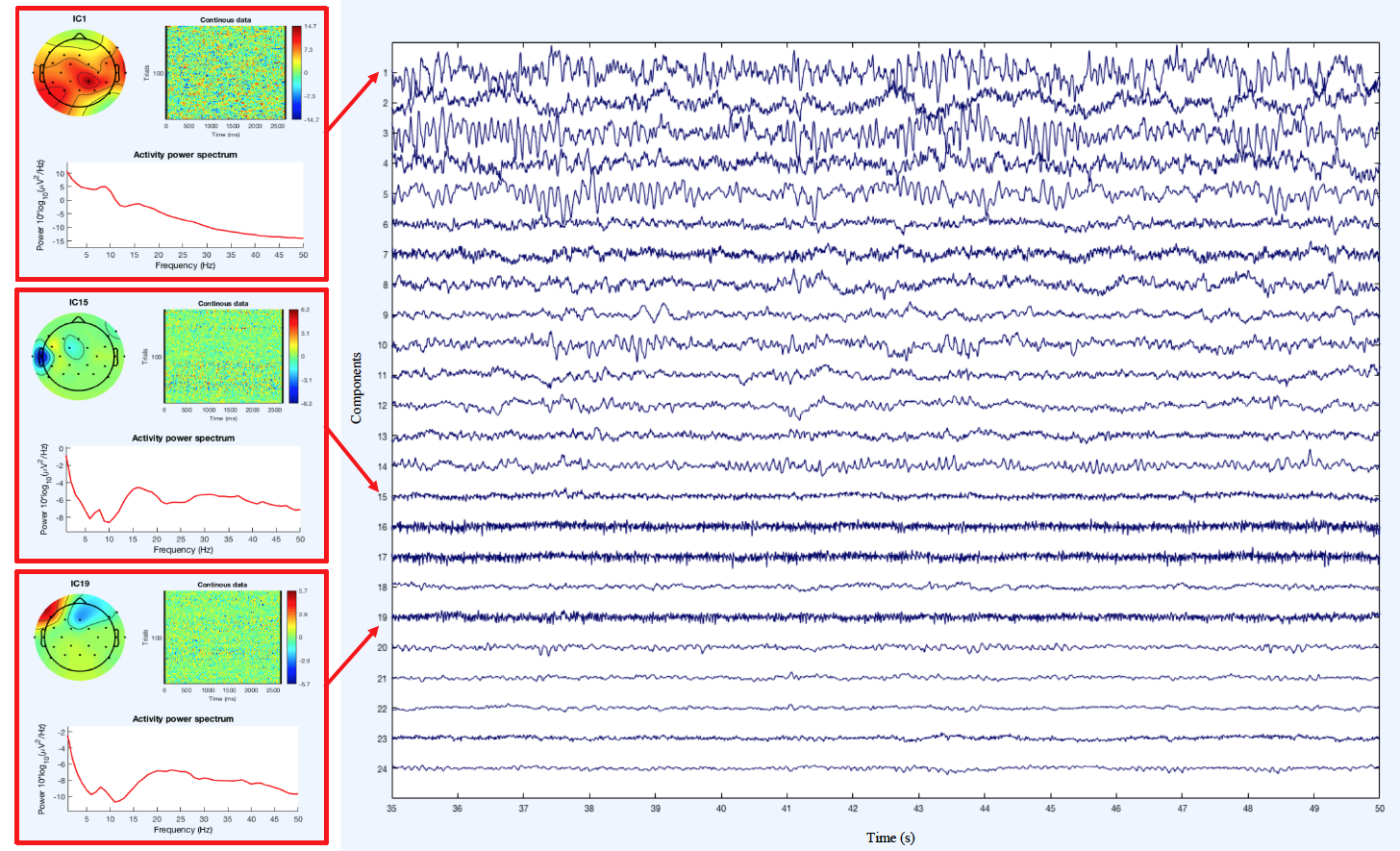}
        \caption{ICA analysis.}
        \label{fig:eeg_ts_ica}
    \end{subfigure}
    \hfill
    \caption{Superposition plot \protect\subref{fig:eeg_ts} of an acquired EEG time series from a subject with autism spectrum disorder, pre-ASR (red) and post-ASR (blue); ICA of the time series \protect\subref{fig:eeg_ts_ica} resulting in 24 independent components (ICs). To the left are 3 ICs with respective scalp topographies and activity power spectra. Component IC1 (top) indicates theta, alpha and beta band activity over temporal parietal regions. Components IC15 (middle) and IC19 (bottom) indicate muscle and ocular movement artifact, respectively.}
    \label{fig:ee_ts}
\end{figure}

Subsequent to acquisition, the raw EEG data output will be imported into the open-source MATLAB toolbox: EEGLAB \cite{delorme2004eeglab}.
The following preprocessing pipeline (see Fig. \ref{fig:pp_pipeline}) will be applied:
% ==========
\begin{enumerate}[itemsep=-1mm]
\item Remove low frequency baseline drift with a 1 Hz high-pass filter.
\item Remove 50-60 Hz AC line noise by applying the \texttt{CleanLine} plugin.
\item Clean continuous raw data using the \texttt{clean\_rawdata} plugin \cite{mullen2015real}.
    \begin{enumerate}[itemsep=-1mm]
        \item It first performs bad channel rejection based on two criteria: (1) channels that have flat signals longer than 5 seconds and (2) channels poorly correlated with adjacent channels.
        \item It then applies artifact subspace reconstruction (ASR) -- an algorithm that removes non-stationary, high variance signals from the EEG then uses calibration data (1 min sections of clean EEG) to reconstruct the missing data using a spatial mixing matrix (see Fig.~\ref{fig:eeg_ts}).
    \end{enumerate}
\item Interpolate removed channels.
\item Re-reference channels to average reference.
\item Separate non-brain artifacts from the EEG recording via EEGLAB's Independent Component Analysis (ICA).
\footnote{Details regarding performing ICA in EEGLAB can be found here: Swartz Center for Computational Neuroscience (2018, September 19). Chapter 09: Decomposing Data Using ICA. EEGLAB Wiki. \url{https://sccn.ucsd.edu/wiki/Chapter_09:_Decomposing_Data_Using_ICA}.}
Briefly, ICA involves the linear decomposition of the aggregate channel activity into a series of independent components that are spatially filtered from the recorded EEG time series.
Components representing eye, cardiac, and muscle artifact are removed and components representing genuine brain activity are retained.
Fig.~\ref{fig:eeg_ts_ica} illustrates the ICA decomposition results in EEGLAB.
\end{enumerate}
% ==========
% ==========

\subsection{EEG Measures during Joint Attention}
We have recently employed preliminary feature analysis on acquired raw EEG data from the work of \cite{jaime2016brief}, wherein the EEG was recorded from adolescents with ASD (N=24) and typically developing adolescents (N=28) while they watched a series of 30-second joint attention eliciting video clips.
First, we applied the pre-processing pipeline (see Section \ref{sec:eeg-acq-pre-prop}) on the the raw EEG time series to remove noisy channels and data segments containing movement and ocular artifacts from the EEG data.
The pre-processed data was then classified using our EEG Analysis Pipeline implemented in Python \cite{jayarathna2018bigdata}.

Joint attention is the ability to socially coordinate visual attention, share a point of view with another person, and process self and other related information.
Hence the data retrieval was performed while making the subjects watch video clips that would help in examining joint attention.
There were a total of 12 videos each of which was 30 seconds.
About one second gap was provided between each video.
Both the EEG and Eye movement were collected while the participants watched the video.
A total of 34 participants EEG data was used in this paper after the preprocessing step.

\subsubsection{Evaluation and Results}
\begin{figure}[hbt!]
    \centering
    \includegraphics[width=.5\textwidth]{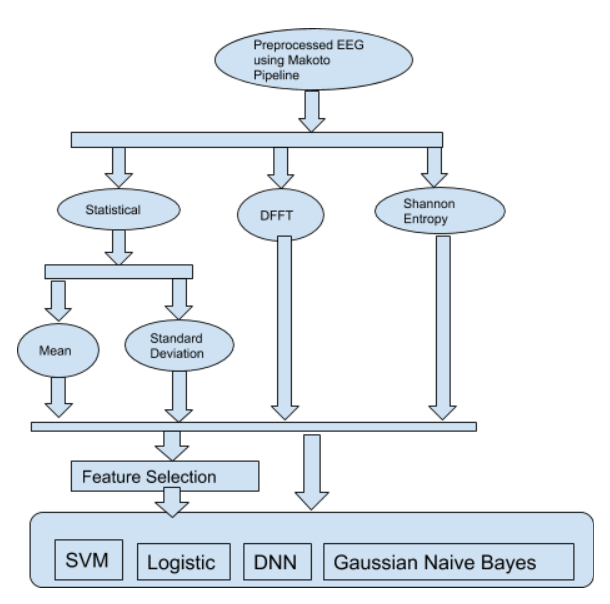}
    \caption{EEG Processing Pipeline for Study 1.}
    \label{fig:eeg_pipeline}
\end{figure}

There are many ways to extract feature from EEG data.
Entropies, wavelets, FFT and various other statistical methods are commonly computed features \cite{al2014methods}.
In this work we use Statistical and Entropy values.
Statistical features comprises of Mean, Standard Deviation, and combined mean and standard deviation of the filtered data.
For the feature analysis, we used statistical and entropy values including mean, standard deviation, and combined mean and standard deviation on the pre-processed data.
Entropy is computed by using Shannon entropy function \cite{lin1991divergence}, which is the average rate at which information is produced by a stochastic source of data given by:
% ==========
\begin{equation}
    H_e=-\sum p_{i} \log_2 p_{i}
\end{equation}
% ==========
Mean function takes in a 2D matrix consisting of the EEG signal of a person and returns a feature vector with mean values for each channel over windows of signal.
For the mean, each of the 128 channels were computed.
For each subject a feature vector consisting of mean of single channel was created.
So the mean function takes in a 2D matrix consisting of the EEG signal of a person and returns a feature vector with mean values for each channel.
For the standard deviation, each of the 128 channels were computed.
For each subject a feature vector consisting of mean of single channel was created. 
So the deviation function takes in a 2D matrix consisting of the EEG signal of a person and returns a feature vector with standard deviation values for each channel (see Fig. \ref{fig:eeg_pipeline}).
% ==========
% ==========
Several models including SVM, Logistic, Deep Neural Network (DNN), and Gaussian Naive Bayes were developed for classification.

\begin{figure}[hbt!]
    \centering
    \includegraphics[width=.5\textwidth]{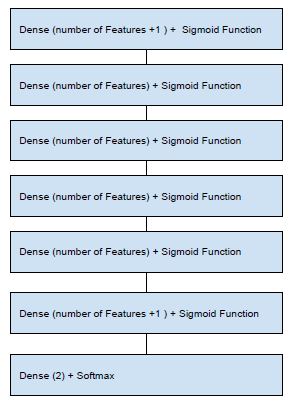}
    \caption{Layers of the Deep Neural Network.}
    \label{fig:dnn_layers}
\end{figure}

For the deep neural network, five hidden layers with sigmoid activation function is used (see Fig. \ref{fig:dnn_layers}).
% ==========
% ==========
Categorical cross entropy was used to calculate loss and ADAmax was used as the optimizer \cite{freivalds2017improving}.
We captured three different feature set; entropy features, FFT and statistical features.
We also calculate mean, and standard deviation.
In total there are 4 different features from EEG and 4 different models for each type of classifier and, overall there are 16 different model variations based on the features (4 feature set x 4 classifiers).
For each feature there are three models for each algorithm, two models using Feature Selection and the third one without using any feature selection.
For Feature selection PCA and sequential feature selection is used.

Table \ref{tab:classifiers-s1} presents an analysis and comparison of EEG data.
% ==========
\begin{table}[hbt!]
    \centering
    \caption{Classification Accuracy of EEG during Joint Attention Study}
    \label{tab:classifiers-s1}
    \begin{tabular}{ccccl}
        \toprule
        Classifier              & Entropy   & FFT           & Mean  & Std   \\
        \midrule
        Gaussian Naive Bayes    & 0.26      & 0.53          & 0.55  & 0.55  \\
        Logistic Regression     & 0.11      & \textbf{0.78} & 0.58  & 0.50  \\
        SVM                     & 0.11      & 0.56          & 0.55  & 0.55  \\
        DNN                     & 0.20      & 0.52          & 0.58  & 0.45  \\
        \bottomrule
    \end{tabular}
\end{table}
% ==========
Note that two models were created for each model with only EEG and combined data by using PCA and without using PCA (e.g. SVM with PCA and SVM without PCA).
For some models, with PCA did better while for some without PCA did better.
For example, DNN almost always without using PCA did worse because of the curse of dimensionality.
The highest performing SVM with about 56\% accuracy was using Shannon Entropy with all the features without PCA.
The highest performing Logistic regression with 78\% accuracy was using a combination of EEG Standard Deviation and eye data without PCA.
SVM, Logistic Regression, and Gaussian Naive Bayes do better without PCA which means that with PCA it loses data points that these models find useful.
This is interesting because PCA is supposed to find the most discriminant features and remove redundant or noisy features.
And this is supposed to help machine learning models produce better results.
For SVM most models with PCA did better except the highest performing model.
This might mean that the Entropy data is more linear than the other datasets.
For DNN the curse of dimensionality is obvious.
Whereas for Gaussian Naive Bayes all the high performing models did not use PCA except the one with the combination of EEG mean.
This is an exception and must be due to the nature of the EEG mean data.
But in general case Naive Bayes does better without PCA.
This might be due to the fact that probabilistic models are able to make sense of higher dimensional dataset much easier than other models like DNN.
Then with using Sequential Feature Selection algorithm almost all the models performed better than either PCA or no Feature Selection.

In this study we have used PCA, and Sequential Feature Selection algorithm.
There are other Feature Selection algorithms like Genetic algorithm, Particle Swarm Optimization, and TWIST which can be compared to find features to optimize the performance of the models.
Also, this will tell us which feature selection algorithm will work better for the combined data sets. 
Gaussian Naive Bayes with some of the features had perfect score.
But we need to reproduce this result with large number of participants to be able to use this in a clinical setting.
Current number of 34 participants is too low to confirm our results.
However, this is a first step towards developing an optimal Autism Diagnosis system.

\subsection{EEG Coherence during Live Social Interaction}
The notion that social brain system FC may be a useful index of social impairment is suggested by both the literature \cite{mundy2016autism,jaime2016brief} and by our preliminary findings obtained from our pilot sample composed of individuals between the ages of 5 and 17 years who completed an ADOS-2 assessment while we simultaneously recorded their EEG (see Table \ref{tab:ados}).
% ==========
\begin{table}[hbt!]
    \centering
    \caption{ADOS-2 Score of the ASD Vs TD}
    \label{tab:ados}
    \begin{tabular}{ccccl}
        \toprule
        Participant & Sex   & Age   & ADOS-2    & Diagnosis \\
        \midrule
        2           & M     & 10    & 19        & ASD       \\
        4           & M     & 17    & 12        & ASD       \\
        11          & M     & 6     & 11        & ASD       \\
        12          & M     & 16    & 16        & ASD       \\
        13          & F     & 11    & 16        & ASD       \\
        15          & F     & 10    & 7         & ASD       \\
        18          & M     & 5     & 20        & ASD       \\
        20          & M     & 15    & 9         & ASD       \\
        \midrule
        5           & M     & 11    & 5         & TD        \\
        7           & F     & 9     & 0         & TD        \\
        8           & F     & 6     & 5         & TD        \\
        14          & F     & 16    & 0         & TD        \\
        16          & M     & 8     & 4         & TD        \\
        17          & F     & 6     & 0         & TD        \\
        19          & M     & 15    & 2         & TD        \\
        21          & M     & 6     & 4         & TD        \\
        22          & F     & 8     & 0         & TD        \\
        \bottomrule
    \end{tabular}
\end{table}
% ==========
Despite a small sample size (ASD = 8; TD = 9), our preliminary results indicate a trending negative association between right hemisphere delta and theta band EEG coherence and level of social symptom severity (according to the ADOS-2 algorithm scoring) in children with ASD (see Table \ref{tab:ados} below), but not in our pilot sample of typically developing (TD) children.
Our preliminary results paint a conceptual picture that is in line with our prior work evaluating EEG coherence during joint social attention perception in ASD \cite{jaime2016brief}, that there are diagnostic group differences in the association between right hemisphere frontal--temporal--parietal FC and standardized measures of social functioning.
Such diagnostic group differences in FC association patterns reflect a tendency for children with impaired social capacity to have idiosyncratic patterns of social brain system functional organization relative to typical neurodevelopment.
Thus, EEG measures of social brain system FC acquired during live social interaction shows promise as a candidate non-invasive biomarker of early emerging aberrant social neurocognitive dysfunction in ASD. 

\subsubsection{EEG Measures of Functional Connectivity}
We first extracted 180-second epochs beginning from the middle one-third portion of each subject's pre-processed EEG time series in order to calculate a functional connectivity (FC) measure of the engaged social brain system.
With each subject's epoched EEG time series treated as a discrete-time signal $u=x_i(t)$ for EEG channel $i$, we used EEG coherence as a variable of FC.
EEG coherence, or normalized magnitude-squared coherence (MSC), $C_{uv}^2(\omega)$, is a statistical estimate of the amount of phase synchrony between two EEG time series $u$ and $v$:
% ==========
\begin{equation}
    C_{uv}^2(\omega) = {|\phi_{uv}(\omega)|^{2}}/{\left(\phi_{uu}(\omega)\phi_{vv}(\omega)\right)}
\end{equation}
% ==========
where the squared magnitude of the cross spectrum density $|\phi_{uv}(\omega)|^{2}$ (a measure of co-variance) between the two signals $u$ and $v$ at a given frequency $\omega$, is normalized by the Power Spectral Densities (PSDs) (variance) of each channel $\phi_{uu}$ and $\phi_{vv}$ so that $0\!\leq C_{uv}^2(\omega) \leq 1$.
Higher values represent greater synchronous activity between distinct channels whereas lower values represent reduced or non-synchronous activity \cite{nunez2006theoretical}.
Coherence is a function of frequency: to compute a single similarity metric between a pair of signals, we integrate over frequency to obtain total power (or variance in a statistical sense) $P_{ij}$ where $T$ is the extent of frequency components sampled.
% ==========
\begin{equation}
    P_{ij} = \frac{1}{T}\int_{0}^{T}{C_{uv}^{2}(\omega)}
\end{equation}
% ==========
The MSC of a signal which itself produces no variance (in the statistical sense) and hence $P_{ii}\!=\!1$, gives a convenient, normalized metric of similarity.

Accordingly, intra-hemispheric MSC between electrode positions that are spatially collocated over areas comprising the social brain system \cite{saxe2006uniquely,adolphs2009social} were examined.
Electrode pairs were selected based on \cite{homan1987cerebral}’s electrode placement correlates of cortical location.
Using the international 10/20 placement system \cite{klem1999ten}, the following electrodes were selected: \textbf{F7, F8, T7, T8, TP9, TP10, P7, P8, C3, C4}.
A mean hemispheric MSC score was calculated across all left and all right electrodes (10 electrode pairs per hemisphere) representing left and right social brain system FC, respectively.

\subsubsection{Evaluation and Results}
\label{classification-results-p1}
We generated five feature sets categorized according to the frequency bands: 1) delta, 2) theta, 3) alpha, 4) beta and 5) gamma with each set representing the amplitude and power of the signal from each electrode.
These feature sets were entered into 43 different classifiers yielding precision rates, recall rates, F1 scores, and percent accuracy.
We identified six the top performing classifiers: RandomForest, Logistic, Bagging, JRip, LMT and AdaBoostM1.

The six top performing classifiers for the 5-band feature set are listed in Table \ref{tab:classifiers}.
These results were calculated based on features from all electrodes.
% ==========
\begin{table}[hbt!]
    \centering
    \caption{Classification of EEG during ADOS-2}
    \label{tab:classifiers}
    \begin{tabular}{ccccl}
        \toprule
        Classifier & Precision & Recall & F1 & Accuracy\\
        \midrule
        RandomForest & 0.98 & 0.98 & 0.98 & \textbf{98.00 \%} \\
        Logistic & 0.96 & 0.96 & 0.96 & 96.63 \% \\
        Bagging & 0.95 & 0.95 & 0.95 & 95.66 \% \\
        JRip & 0.98 & 0.98 & 0.98 & \textbf{98.06 \%} \\
        LMT & 0.95 & 0.95 & 0.95 & 95.79 \% \\
        AdaBoostM1 & 0.92 & 0.92 & 0.92 & 92.14 \% \\
        \bottomrule
    \end{tabular}
\end{table}
% ==========
The JRip classifier yielded the highest percent accuracy with 98.06\% indicating that a 5-band feature set collected during an ADOS-2 test classifies a diagnosis of ASD with greater than 90\% accuracy.
From these six classifiers, the AdaBoostM1 classifier yielded the lowest percent accuracy at 92.14\%.

We also conducted an evaluation by selecting only \textbf{F7, F8, T7, T8, TP9, TP10, P7, P8, C3, C4} electrodes based on \cite{homan1987cerebral}’s electrode placement correlates of cortical location.
The results of this evaluation are listed in Table \ref{tab:classifiers-selected-features}.
% ==========
\begin{table}[hbt!]
\centering
  \caption{Classification of EEG during ADOS-2 with Selected Features}
  \label{tab:classifiers-selected-features}
  \begin{tabular}{ccccl}
    \toprule
    Classifier & Precision & Recall & F1 & Accuracy\\
    \midrule
    RandomForest & 0.97 & 0.97 & 0.97 & \textbf{97.04 \%} \\
    Logistic & 0.84 & 0.84 & 0.84 & 84.72 \% \\
    Bagging & 0.95 & 0.95 & 0.95 & \textbf{95.50 \%} \\
    JRip & 0.94 & 0.94 & 0.94 & 94.57 \% \\
    LMT & 0.83 & 0.82 & 0.82 & 82.94 \% \\
    AdaBoostM1 & 0.80 & 0.79 & 0.79 & 79.75 \% \\
  \bottomrule
\end{tabular}
\end{table}
% ==========
When comparing the results, it was observed that the the RandomForest classifier yielded the highest percent accuracy with 97.04\%.
The AdaBoostM1 classifier yielded the lowest percent accuracy at 79.75\%.

\subsection{Temporal Relationship between ASD and Brain Activity}
In this analysis, we derive power spectrums for each electrode through frequency band decomposition and through wavelet transforms relative to a baseline, and generate two sets of training data that captures long-term and short-term trends respectively.
We utilize machine learning models to predict the ASD diagnosis and the ADOS-2 scores, which provide an estimate for the presence of such trends.
When evaluating short-term dependencies, we obtain a maximum of 56\% accuracy of classification through linear models.
Non-linear models provide a classification above 92\% accuracy, and predicted ADOS-2 scores within an RMSE of 4.
We use a CNN model to evaluate the long-term trends, and obtain a classification accuracy above 90\%.

\subsubsection{Feature Extraction and Analysis}
We analyzed the signal generated from the pre-processing stage in two approaches.

\begin{itemize}
    \item Method 1: Frequency Band Decomposition
    \item Method 2: Wavelet Transform
\end{itemize}

Both approaches were evaluated independent of each other via multiple machine learning models to obtain conclusive results.
% ===================================================================
\vspace{5pt}
\paragraph{Frequency Band Decomposition}
Time series data of each electrode was passed into a function that decomposed it into five distinct time series, each containing only the data within its corresponding frequency band.
This decomposition was performed using Butterworth filters in the order of 5 (n = 5), where mathematical equation is given in Equation \ref{eq:butterworth}.

\begin{equation}
\label{eq:butterworth}
|H(j\omega)| = 1/\sqrt{1+\epsilon^2(\omega/\omega_p)^{2n}}
\end{equation}
Here, $\omega_p$ is the cutoff angular velocity, and $\epsilon$ is the maximum band pass gain.
We extracted time series data for five frequency bands $\delta$, $\theta$, $\alpha$, $\beta$ and $\gamma$, with each band corresponding to the frequency ranges shown in Table \ref{tab:freq:bands}.

\begin{table}[hbt!]
    \centering
    \caption{Frequency bands used for decomposition}
    \label{tab:freq:bands}
    \begin{tabular}{ccc}
    \toprule
    Band        & Lo (Hz)   & Hi (Hz)\\
    \midrule
    $\delta$    & 0.1       & 4.0\\
    $\theta$    & 4.0       & 7.5\\
    $\alpha$    & 7.5       & 12.0\\
    $\beta$     & 12.0      & 30.0\\
    $\gamma$    & 30.0      & 100.0\\
    \bottomrule
    \end{tabular}
\end{table}

\begin{figure}[hbt!]
    \centering
    \begin{subfigure}[b]{0.45\textwidth}
        \includegraphics[width=\textwidth]{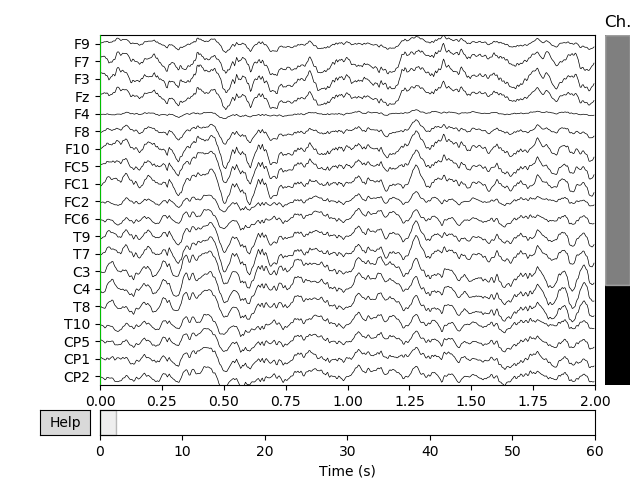}
        \caption{Electrode readings after preprocessing}
        \label{fig:bandpass:before}
    \end{subfigure}
    \begin{subfigure}[b]{0.45\textwidth}
        \includegraphics[width=\textwidth]{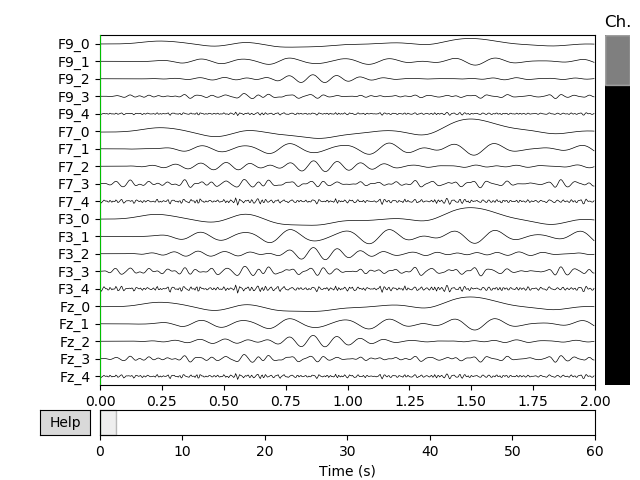}
        \caption{Electrode readings after band-pass filtering}
        \label{fig:bandpass:after}
    \end{subfigure}
    \caption{EEG Pre-processing and Band Pass Filtering}
    \label{fig:eeg:pre-processing}
\end{figure}

Each series represents data within one frequency band of an electrode.
For instance, F9\_0 represents the $\delta$ frequency band of the F9 electrode, and F9\_1 represents the $\theta$ frequency band of that electrode (see Fig. \ref{fig:eeg:pre-processing}).

An algorithm was devised to calculate the power matrix for a time series as shown in Algorithm \ref{alg:pow-matrix}.
% ===================================================================
\begin{algorithm}[hbt!]
\centering
\caption{Power Matrix of an Electrode}
\label{alg:pow-matrix}
\begin{algorithmic}[1]{
    \State $f \gets sampling\_freq$
    \Function{PowerMatrix}{$B$, $S$, $W$, $E$}
        \State $F \gets BandPass(B, S)$
        \State $I \gets |B|$
        \State $J \gets |S| / S / f$
        \State $M \gets array[I][J]$
        \ForAll{$i \gets 0, .. (I-1)$}
            \ForAll{$j \gets 0, .. (J-1)$}
                \State $M[i][j] \gets P(F[i], W, E, j, f)$
            \EndFor
        \EndFor
        \\\Return $M$
    \EndFunction
}\end{algorithmic}
\end{algorithm}
Given a set of frequency bands $B$, a time series $S$ with $n$ samples recorded at a sampling frequency of $f$, a window size of $W$ seconds and a step size of $E$ seconds, it generates the power matrix by decomposing the signal $S$ into $|B|$ signals corresponding to each band in $B$, stepping through each decomposed signal in strides of size $f \times E$, and calculating the power within a window of size $f \times W$.
Thus, a $(|B| \times (n/f\times s) )$ matrix was generated for each electrode.
The function P used in this Algorithm is described in Equation \ref{eq:pow}.
It calculates the power of a signal $S$ within a window $W$ for a step size $E$, given the current step $j$ and sampling frequency $f$.
\begin{equation}
\label{eq:pow}
P(S,W,E,j,f) = \frac{1}{W}\sum_{k=0}^{fW-1} |S[Efj + k]|^2
\end{equation}
% ===================================================================

We initialized the window size and step size to $W = 5$ and $S = 2$ respectively, and generated power matrices for all electrodes of each participant at all epochs.
% ===================================================================
\vspace{5pt}
\paragraph{Wavelet Transform}

Wavelet transforms were used as an improvement over the former approach to compensate for the loss of resolution when using a fixed-size window to calculate power matrices.
The wavelet transform function is described in Equation \ref{eq:wt}, 
\begin{equation}
    \label{eq:wt}
    X(a,b) = \frac{1}{\sqrt{a}} \int_{-\infty}^{+\infty} \psi \bigg(\frac{t-b}{a}\bigg) x(t) dt
\end{equation}
where $a$ and $b$ corresponds to the scale and translation of the wavelet signal $\psi$, which is convolved over the source signal $x(t)$.

A Complex Morlet Wavelet (cmor) with a center frequency $f_c = 1 Hz$ and a bandwidth of $f_b = 1.5 Hz$ was used for this task.
We generated wavelet transforms for each signal S at X different scales (X = 150) starting from $1 \times (f/2) = 125 Hz$ down to $1 \times (f/2) / 150 \approx 0.8 Hz$
and generated a ($X \times |S|$) matrix for each electrode.
The resulting matrix was scaled down to $(X \times X)$ using a max aggregator function, to reduce computational complexity.
All matrix coefficients were squared to obtain their power equivalent.

Next, each power matrix was referenced to the mean signal strength of the electrode's baseline.
When calculating the baseline value for an electrode, a wavelet transform was performed on the BASELINE epoch of that electrode and the resulting matrix was reduced to a column vector using a mean aggregator.
This column vector was subtracted from each column of the electrode power matrix to obtain the final matrix.

\begin{figure}[hbt!]
\vspace{-10pt}
    \centering
    \begin{subfigure}[b]{0.3\textwidth}
        \includegraphics[width=\textwidth,trim={80pt 0 110pt 15pt},clip]{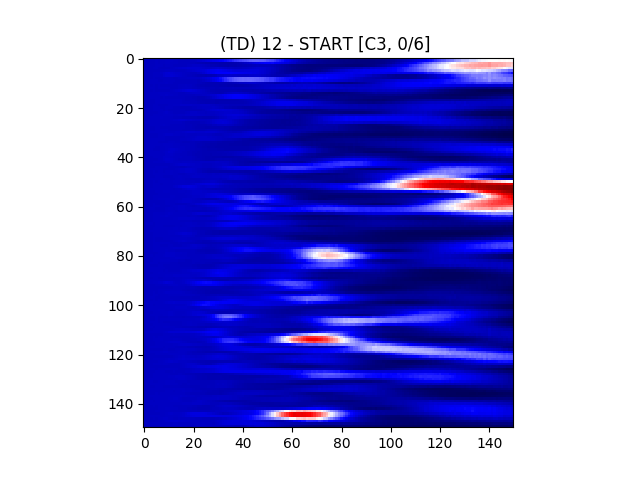}
        \caption{Diagnosis - TD}
        \label{subfig:wt-td}
    \end{subfigure}
    \begin{subfigure}[b]{0.3\textwidth}
        \includegraphics[width=\textwidth,trim={80pt 0 110pt 15pt},clip]{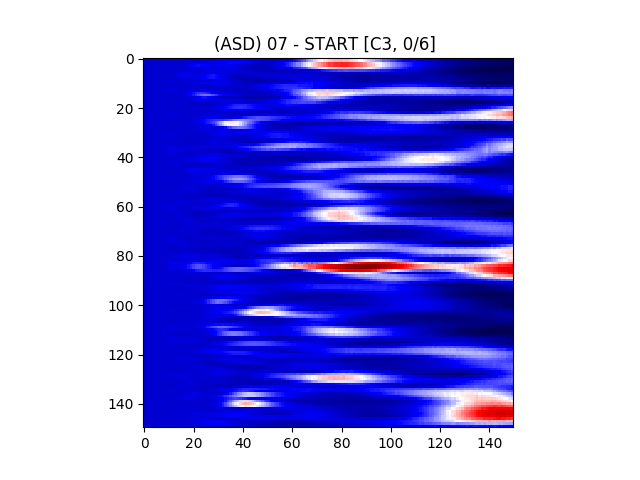}
        \caption{Diagnosis - ASD}
        \label{subfig:wt-asd}
    \end{subfigure}
    \caption{Wavelet Transforms for electrode FC1}
\end{figure}

Figure \ref{subfig:wt-td} illustrates the power matrix of a sample electrode for a TD (typically developing) subject.
All values were normalized to the (0 - 255) range for illustration purposes.
The X axis represents time scaled by a factor of $4x$, and the Y axis represents the scale of the reference wavelet used, which relates to the frequency $f(s)$ as given in Equation \ref{eq:scale2freq}
\begin{equation}
    f(s) = f_c \times \frac{f}{s}
    \label{eq:scale2freq}
\end{equation}
where $f_c$ is the center frequency of the wavelet, f is the sampling frequency, and s is the scale.
The color at each $(x, y)$ coordinate corresponds to the signal power at that time and scale.
Darker shades represent lower power, while lighter shades represent higher power.
Figure \ref{subfig:wt-asd} shows the illustration of a wavelet transform performed on the FC1 electrode, but for an ASD subject.
The axes and colors follow the same convention as Figure \ref{subfig:wt-td}.
Both images are referenced to the mean signal strength of the corresponding electrode as observed in the BASELINE epoch.

\begin{figure}[hbt!]
    \centering
    \begin{subfigure}[b]{0.2\textwidth}
        \includegraphics[width=\textwidth,trim={80pt 0 110pt 15pt},clip]{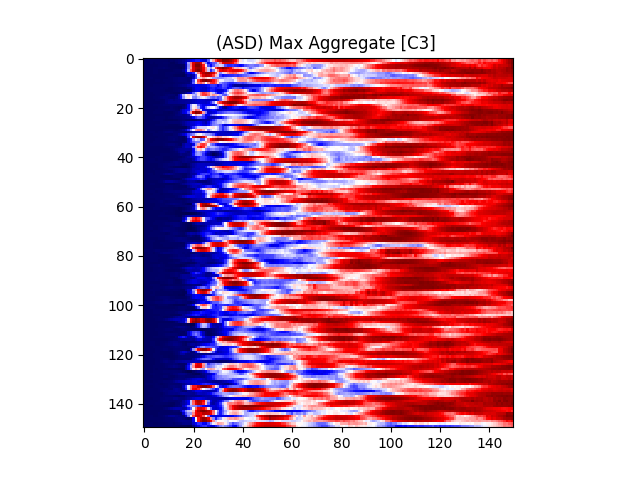}
        \caption{C3}
        \label{subfig:wt-asd-agg-1}
    \end{subfigure}
    \begin{subfigure}[b]{0.2\textwidth}
        \includegraphics[width=\textwidth,trim={80pt 0 110pt 15pt},clip]{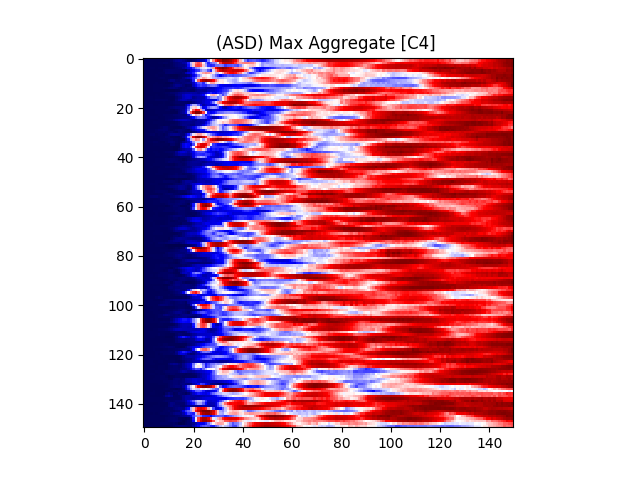}
        \caption{C4}
        \label{subfig:wt-asd-agg-2}
    \end{subfigure}
    \begin{subfigure}[b]{0.2\textwidth}
        \includegraphics[width=\textwidth,trim={80pt 0 110pt 15pt},clip]{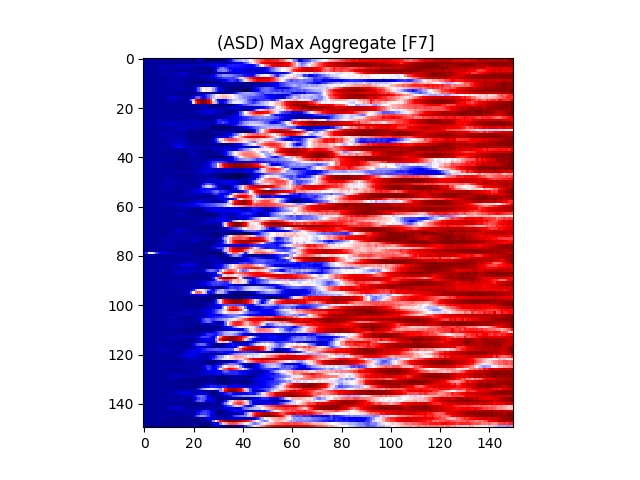}
        \caption{F7}
        \label{subfig:wt-asd-agg-3}
    \end{subfigure}
    \begin{subfigure}[b]{0.2\textwidth}
        \includegraphics[width=\textwidth,trim={80pt 0 110pt 15pt},clip]{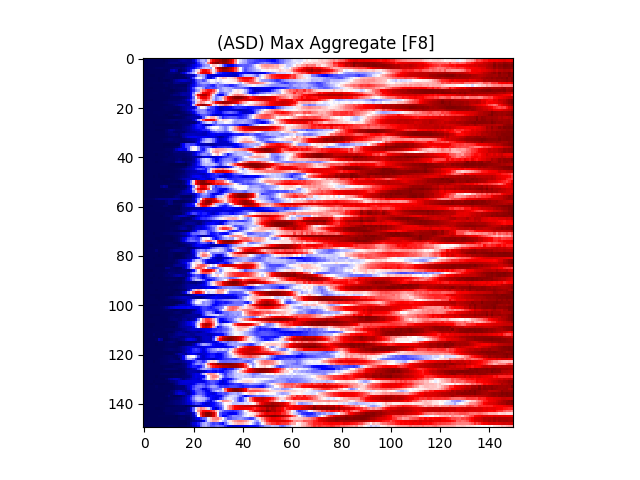}
        \caption{F8}
        \label{subfig:wt-asd-agg-4}
    \end{subfigure}\\
    \begin{subfigure}[b]{0.2\textwidth}
        \includegraphics[width=\textwidth,trim={80pt 0 110pt 15pt},clip]{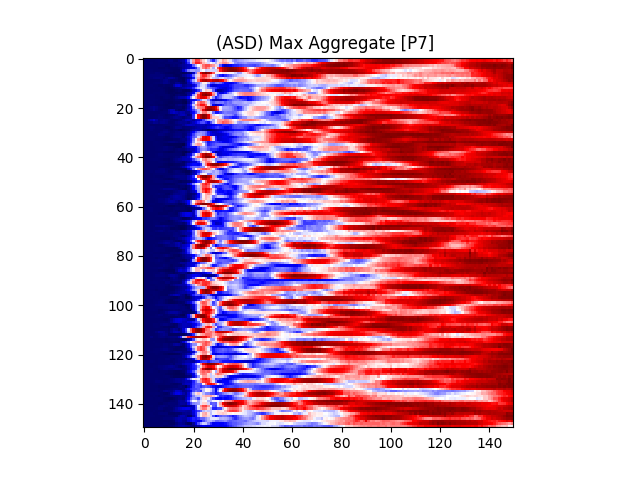}
        \caption{P7}
        \label{subfig:wt-asd-agg-5}
    \end{subfigure}
    \begin{subfigure}[b]{0.2\textwidth}
        \includegraphics[width=\textwidth,trim={80pt 0 110pt 15pt},clip]{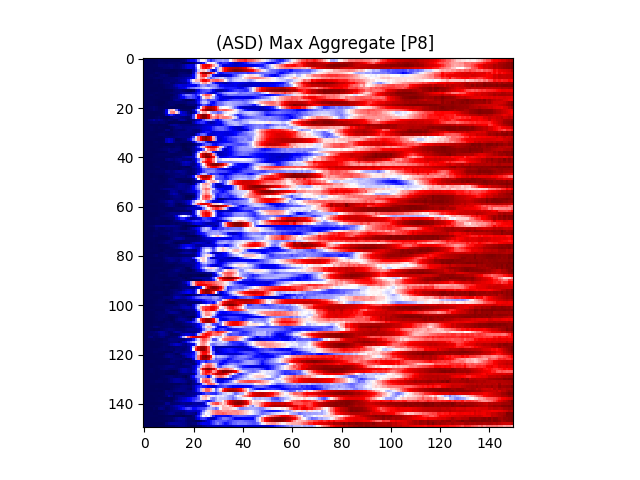}
        \caption{P8}
        \label{subfig:wt-asd-agg-6}
    \end{subfigure}
    \begin{subfigure}[b]{0.2\textwidth}
        \includegraphics[width=\textwidth,trim={80pt 0 110pt 15pt},clip]{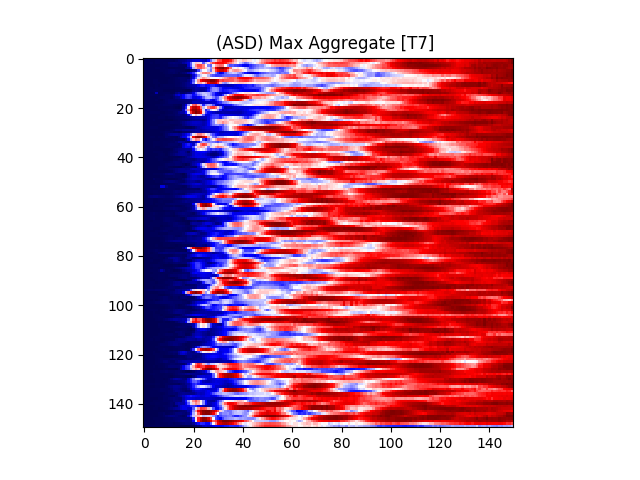}
        \caption{T7}
        \label{subfig:wt-asd-agg-7}
    \end{subfigure}
    \begin{subfigure}[b]{0.2\textwidth}
        \includegraphics[width=\textwidth,trim={80pt 0 110pt 15pt},clip]{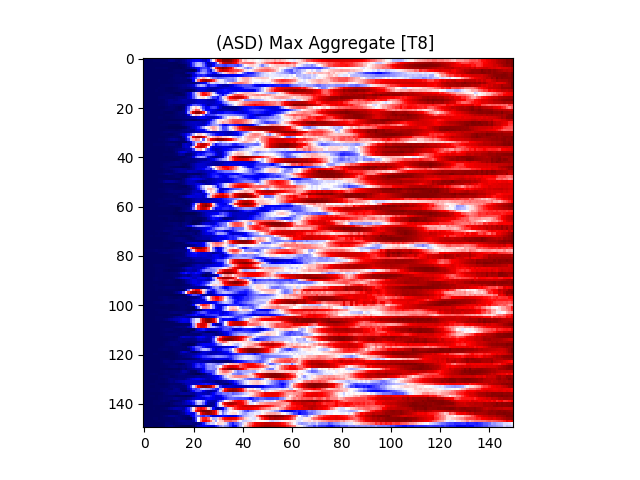}
        \caption{T8}
        \label{subfig:wt-asd-agg-8}
    \end{subfigure}
    \caption{Aggregate Power Spectrums of all ASD Participants}
    \label{fig:wt-asd-agg}
\end{figure}

\begin{figure}[hbt!]
    \centering
    \begin{subfigure}[b]{0.2\textwidth}
        \includegraphics[width=\textwidth,trim={80pt 0 110pt 15pt},clip]{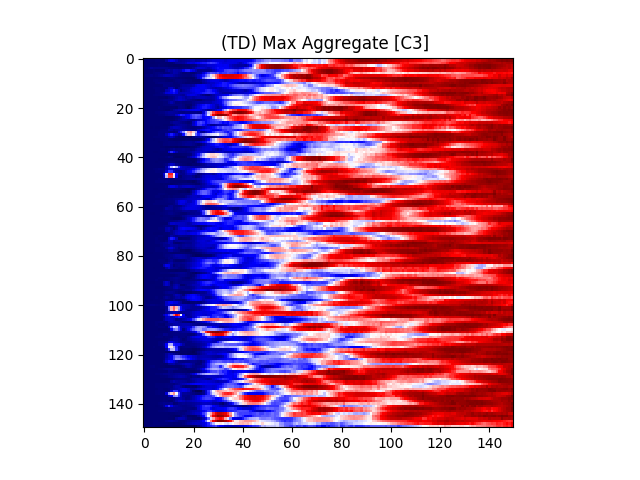}
        \caption{C3}
        \label{subfig:wt-td-agg-1}
    \end{subfigure}
    \begin{subfigure}[b]{0.2\textwidth}
        \includegraphics[width=\textwidth,trim={80pt 0 110pt 15pt},clip]{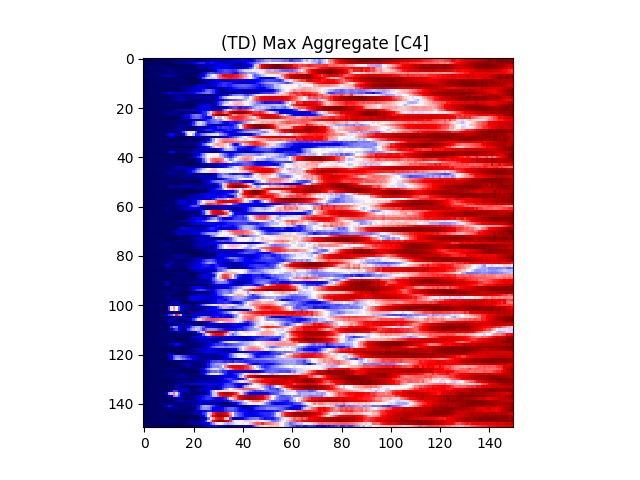}
        \caption{C4}
        \label{subfig:wt-td-agg-2}
    \end{subfigure}
    \begin{subfigure}[b]{0.2\textwidth}
        \includegraphics[width=\textwidth,trim={80pt 0 110pt 15pt},clip]{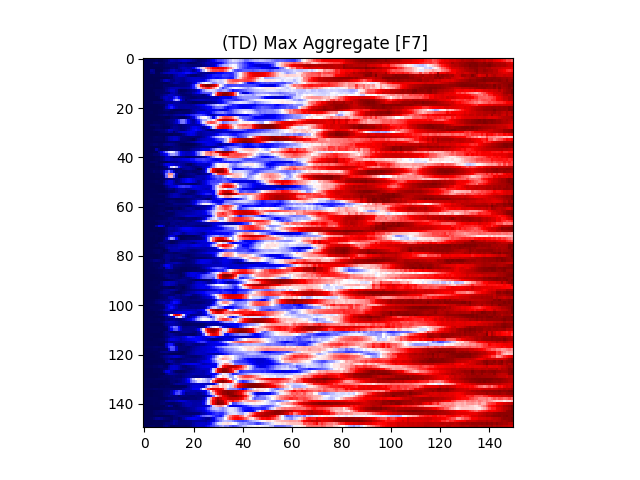}
        \caption{F7}
        \label{subfig:wt-td-agg-3}
    \end{subfigure}
    \begin{subfigure}[b]{0.2\textwidth}
        \includegraphics[width=\textwidth,trim={80pt 0 110pt 15pt},clip]{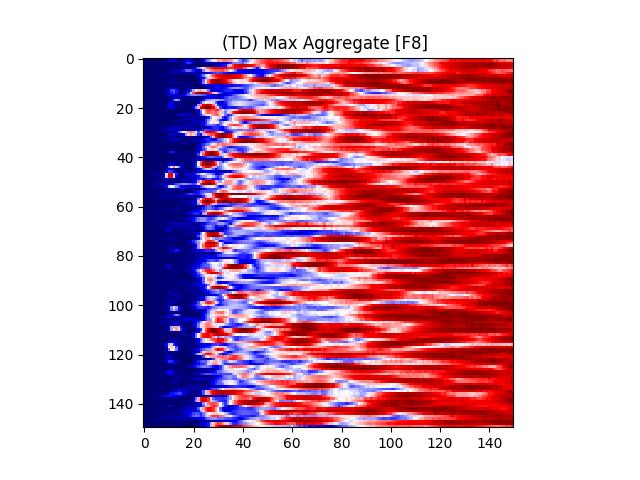}
        \caption{F8}
        \label{subfig:wt-td-agg-4}
    \end{subfigure}\\
    \begin{subfigure}[b]{0.2\textwidth}
        \includegraphics[width=\textwidth,trim={80pt 0 110pt 15pt},clip]{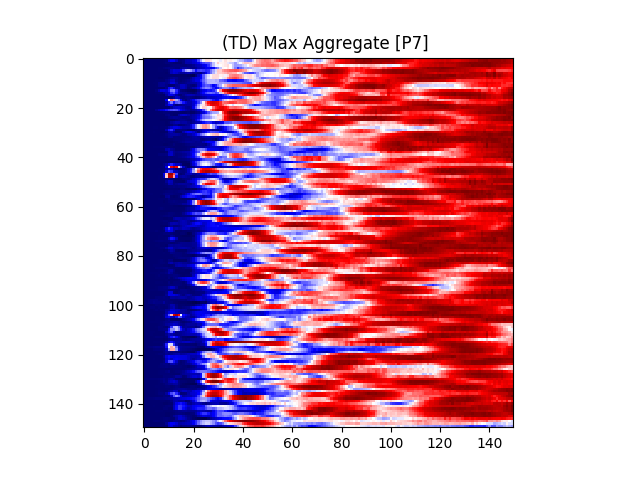}
        \caption{P7}
        \label{subfig:wt-td-agg-5}
    \end{subfigure}
    \begin{subfigure}[b]{0.2\textwidth}
        \includegraphics[width=\textwidth,trim={80pt 0 110pt 15pt},clip]{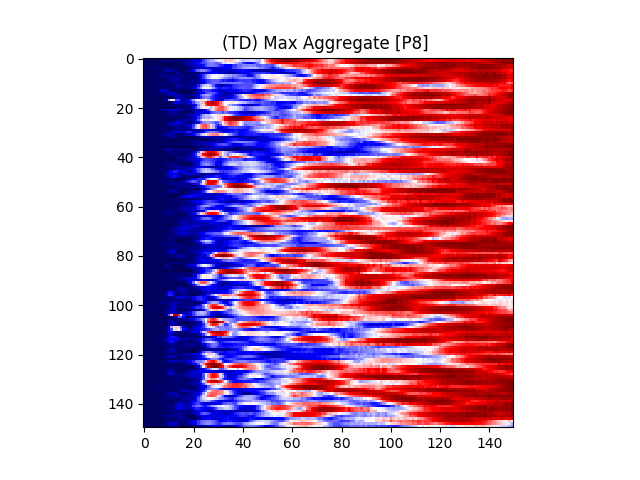}
        \caption{P8}
        \label{subfig:wt-td-agg-6}
    \end{subfigure}
    \begin{subfigure}[b]{0.2\textwidth}
        \includegraphics[width=\textwidth,trim={80pt 0 110pt 15pt},clip]{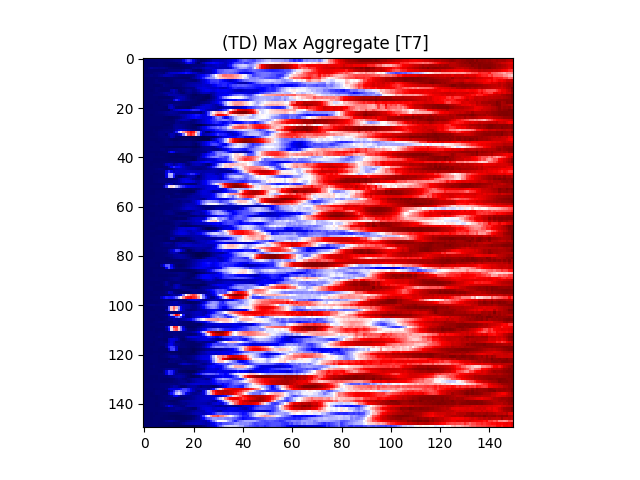}
        \caption{T7}
        \label{subfig:wt-td-agg-7}
    \end{subfigure}
    \begin{subfigure}[b]{0.2\textwidth}
        \includegraphics[width=\textwidth,trim={80pt 0 110pt 15pt},clip]{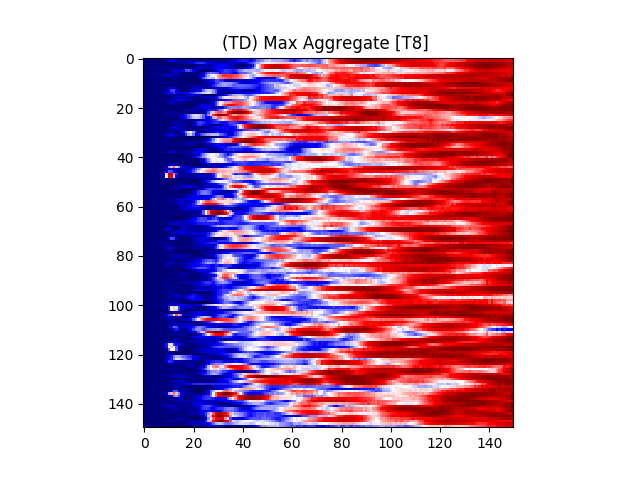}
        \caption{T8}
        \label{subfig:wt-td-agg-8}
    \end{subfigure}
    \caption{Aggregate Power Spectrums of all TD Participants}
    \label{fig:wt-td-agg}
\end{figure}

Figure \ref{fig:wt-asd-agg} provides an illustration for the power spectrums of 8 EEG electrodes, whose color at each $(x,y)$ coordinate corresponds to the maximum power across all epochs and chunks of all ASD participants for that particular coordinate.
A max aggregator function was used to generate this visualization.
% ===================================================================
Similarly, Figure \ref{fig:wt-td-agg} shows illustrations for the same 8 electrodes given in Figure \ref{fig:wt-asd-agg}, but with the color at each $(x,y)$ calculated from TD participants instead of ASD participants.
% ===================================================================
% ===================================================================
% ===================================================================
\subsubsection{Evaluation and Results}
The power matrices obtained through Frequency Band Decomposition and Wavelet Transforms were used to train several machine learning models.
We evaluated both short-term and long-term dependencies between ASD and EEG data to determine the nature of their relationship.
Our evaluation criteria was based on the following objectives.
\begin{itemize}
    \item Providing a clinical diagnosis of ASD for each participant (classification)
    \item Predicting the ADOS-2 score for each participant (regression)
\end{itemize}

\paragraph{Analysis of short-term trends}

Here, our goal was to evaluate whether an accurate diagnosis of ASD could be provided without taking long-term trends into account.
WEKA was used to perform this analysis, and the evaluation measures were obtained by training models using short-term features:
Each power matrix was decomposed into a set of vectors each representing the powers of frequency bands within a window of ($t = 180 / 150 = 1.2 s$).
Electrode vectors belonging to that participant, epoch and timestamp were aggregated into a single vector, and treated as one sample.
We used these samples to train several models using two approaches: 1) using only Homan et al. \cite{homan1987cerebral} electrode placement correlates of cortical locations (F7, F8, T7, T8, TP9, TP10, P7, P8, C3, and C4), and 2) using all electrodes.
Precision, recall, accuracy and F1 statistics were obtained for each model and electrode set through 10-fold cross validation (see Section \ref{classification-results-p1} for results).

Table \ref{tab:ados2-regression} shows the results of the correlation analysis between the predicted and labelled ADOS-2 Scores.
% ===================================================================
\begin{table}[hbt!]
\centering
  \caption{Correlation Analysis for ADOS-2 Score}
  \label{tab:ados2-regression}
  \begin{tabular}{lccc}
    \toprule % =======================================================
    Model               & $r^2$     & MAE       & RMSE              \\
    \midrule % =======================================================
    Random Forest       & 0.8606    & 2.2122    & \textbf{2.9724}   \\
    Linear Regression   & 0.7296    & 2.7978    & 3.4921            \\
    Bagging             & 0.8242    & 1.8986    & \textbf{2.9342}   \\
    REP Tree            & 0.6681    & 2.1463    & 3.8749            \\
    \bottomrule % ====================================================
\end{tabular}
\end{table}
% ===================================================================
Here, Bagging yielded the highest correlation coefficient (r) of 0.9079, with a Root Mean Squared Error (RMSE) of 2.93.
The REP Tree yielded the least correlation coefficient of 0.8174, with a RMSE of 3.87.

\paragraph{Analysis of long-term trends}
The power matrices were also evaluated using a Convolutional Neural Network (CNN) to account for any temporal relationships in the EEG data.
In short-term trend analysis, each column of the power spectrum was taken as an vector regardless of time.
But in long-term trend analysis, the power matrix was used as-is.
However, unlike in short-term analysis, the power matrices of all electrodes belonging to that participant and epoch were treated as independent samples and not aggregated.

\begin{figure}[hbt!]
    \centering
    \includegraphics[width=\textwidth,trim={0 40pt 0 0},clip]{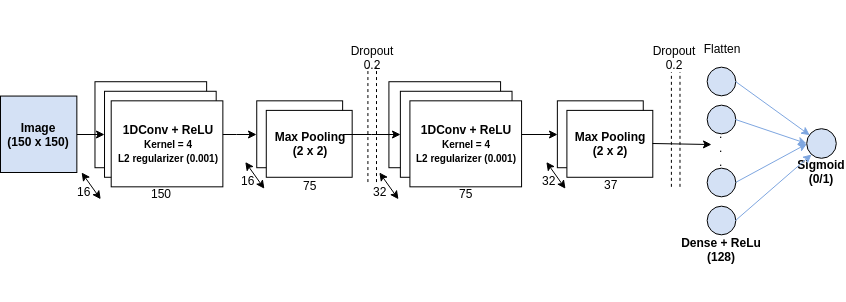}
    \vspace{-20pt}
    \caption{Layers of the CNN model}
    \label{fig:cnn-arch}
    \vspace{-5pt}
\end{figure}

Figure \ref{fig:cnn-arch} shows the structure of the CNN used for this analysis.
The first layer is a 1D Convolution Layer, which uses a kernel of size = 4 to generate (1 x 16) vectors for each convolution.
The output of this layer is passed into a Max Pooling Layer which aggregates adjacent vectors using a Max operator.
The purpose of this layer is to reduce the resolution of the representation learned by the upper layers to enable lower layers to focus on other details.
Next the data passes through a Dropout Layer, which drops 20\% of calculated weights to zero to prevent overfitting.
The output is then passed on to another 1D Convolution Layer, which uses a kernel of size = 4 to generate (1 x 32) vectors for each convolution.
This output is then passed on to another Max Pooling Layer, then to another Dropout Layer, and to a Flatten Layer, which flattens the input matrix into a vector.
This output is passed through a Dense Layer, and finally arrives at a Sigmoid Neuron that performs binary classification based on the input from Dense Layer.
All Convolution Layers were configured with L2 regularizers to avoid overfitting the data.

A stochastic gradient descent (SGD) optimizer was used for training with a learning rate of 0.01, a decay factor of $10^{-6}$ and a momentum of 0.9.
A $3/1$ split was used for train/test data, and the model was trained for 32 epochs.
\begin{figure}[hbt!]
    \centering
    \begin{subfigure}[b]{0.45\textwidth}
        \includegraphics[width=\textwidth]{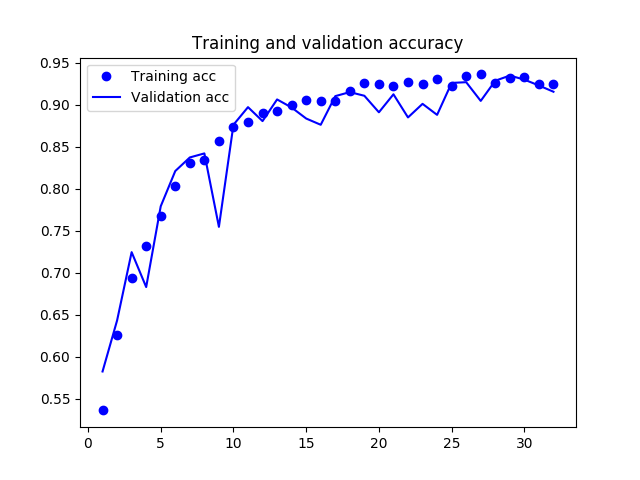}
        \caption{Accuracy}
        \label{fig:cnn-acc}
    \end{subfigure}
    \begin{subfigure}[b]{0.45\textwidth}
        \includegraphics[width=\textwidth]{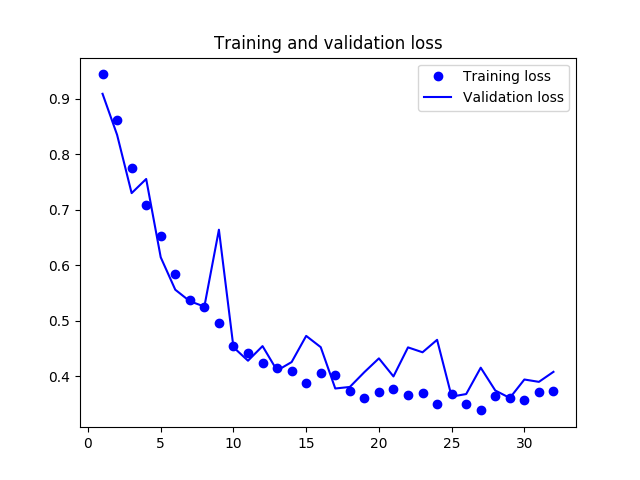}
        \caption{Loss}
        \label{fig:cnn-loss}
    \end{subfigure}
    \caption{Training Progress}
\end{figure}
Figures \ref{fig:cnn-acc} and \ref{fig:cnn-loss} illustrates the training progress of the CNN using the given samples, and the respective change in accuracy and loss metrics across each training epoch.

\subsubsection{Discussion}
Results obtained from short-term and long-term trend analysis shows a high correlation of the EEG data with the human-labeled ASD diagnosis and ADOS-2 scores.
A slight boost in accuracy by moving from electrode set 1 to electrode set 2 was achieved by adding $32 - 10 = 22$ more electrodes to each training sample.
Thus it could be argued that a slight penalty on accuracy is desirable than the added computational complexity from electrode set 2.
The best option of the two depends on the requirements and constraints.

\section{Conclusion and Future Outlook}
Due to its low cost and feasibility, electroencephalography (EEG) shows potential as an effective neurophysiological instrument in the classification of ASD \cite{lenartowicz2014use,snyder2015integration,gloss2016practice}, and there is emerging evidence that -- combined with machine learning approaches -- quantitative measures of EEG can predict ASD with high levels of sensitivity and specificity \cite{bosl2018eeg,grossi2017diagnosis,djemal2017eeg}.
A particular advantage of EEG is its ability to be applied to ecologically valid contexts (i.e., person-to-person social interaction) via wireless solutions thus allowing for the simultaneous acquisition of data from multiple participants in real-world settings. 

To establish proof of concept -- that our classifiers show utility to predict features in line with diagnostic criteria of ASD -- we collect biobehavioral metrics within the context of standardized tasks used in a gold standard assessment of ASD symptomatology: the Autism Diagnostic Observation Schedule Second Edition (ADOS-2) \cite{gotham2007autism}.
The ADOS-2 has been carefully developed to create snapshots of naturalistic social scenarios that can reveal observable features central to ASD (i.e. joint attention, social overtures), thereby allowing us to measure brain activity that are temporally concurrent with these observable ASD features within relatively brief periods.
It is also important to note that we will not use these ADOS-2 tasks as a clinical tool to diagnose participants; rather, we will capitalize on the semi-structured and standardized nature of these social tasks in the ADOS-2 to create a context that engages the social brain system and elicits joint visual attention behavior for acquisition of biobehavioral metrics.
Thus, participants recruited for this study have already received a diagnosis of ASD by a clinical professional prior to enrolling in this study.

Due to its high temporal resolution and feasibility, electroencephalography (EEG) shows potential as an effective neurophysiological instrument in the classification of ASD \cite{lenartowicz2014use, snyder2015integration, gloss2016practice}.
A particular advantage of EEG is its ability to be applied to ecologically valid contexts via wireless solutions that allow for the simultaneous acquisition of data from multiple participants.
This makes EEG an appropriate choice for examining relevant neurophysiological features of ASD in real-world settings \cite{lee2006using}.
Despite these advantages, the majority of EEG research occurs in highly controlled experimental environments, requiring data collected over a large number of trials with minimal head movement.
We will address this deficiency by combining EEG and eye tracker usage.

Early diagnosis is crucial for successful treatment of ASD.
Although progress has been made to accurately diagnose ASD, it is far from ideal \cite{dawson2008early}.
It often requires various subjective measures, behavioral assessments, observations from caretakers over a period of time to correctly diagnose ASD.
Even with this tedious testing often times individuals are misdiagnosed.
However, there remains promise in the development of accurate detection using subjective modalities of EEG, and Eye movements. 

In the future we will obtain two sets of biobehavioral measures representing joint attention: functional integration of neurocognitive networks associated with the social brain (i.e. EEG metrics) and visual behavior (i.e. eye tracking metrics).
With regard to visual behavior, we will collect, analyze, and produce a battery of traditional positional eye movement metrics thought to be potential indicators of joint attention, including number of fixations \cite{JK03}, fixation durations \cite{FJM50, JC76}, and number of regressions \cite{AMYO+14} during naturalistic, dynamic communication tasks.
We will also compare and contrast baseline-related pupil diameter difference measures and our own Index of Pupillary Activity (IPA) for effectiveness of cognitive load measurement \cite{KDNB+18,DKKB+18}.
Specific to the study of ASD, we will develop new gaze measures based on our previously developed gaze transition entropy \cite{KDSK+15}, adapted to the analysis of gaze switching behavior when looking at the communication partner’s face during naturalistic joint attention tasks.
These future directions will therefore create a collection of eye gaze metrics that we consider non-invasive indices of ASD.
Eye gaze measures will be combined with FC outcomes to facilitate comparable variables to determine if there are statistical differences in biobehaviroal measures for children with and without ASD.
For example, we will statistically compare both measures of EEG and metrics such as gaze transition entropy for congruency.
Regression models will then be built to determine which sets of scores best predict diagnostic classification.

% Bibliography
\bibliographystyle{unsrt}

\end{document}